\documentstyle[12pt,aasms4] {article}
\newcommand{\ts}{\thinspace}

\begin{document}

\title{Near-Infrared Spectroscopy of Powerful Radio Galaxies at 
{\bf $z = 2.2-2.6$}}

\author{A. S. Evans\footnote{Present Address: Dept. of Astronomy 105-24,
California
Institute of Technology, Pasadena, CA 91125}$^,$\footnote{Visiting
Astronomer at the United Kingdom Infrared Telescope, which is operated by the
Joint Astronomy Centre on behalf of the U.K. Particle Physics and Astronomy
Research Council.}}
\affil{Institute for Astronomy, 2680 Woodlawn Drive, Honolulu, HI 96822;
ase@astro.caltech.edu}

\begin{abstract}

Near-infrared spectroscopy ($\lambda_{\rm rest}
\sim${\ts}3700--6800{\ts}\AA) of eight high redshift powerful radio
galaxies (HzPRGs) at $z = 2.2-2.6$ is presented.  Strong forbidden
lines and H$\alpha$ emission were detected in all sources; the data
show evidence that the emission lines of the HzPRGs may contribute a
substantial fraction ($\sim$ {\ts}25--98{\ts}\%) of their total
observed $H$- and/or $K$-band light.  Diagnostic emission-line ratios
-- [O III]{\ts}$\lambda$5007{\ts}/{\ts}H$\beta$ vs.  [S
II]{\ts}$\lambda \lambda$6716, 6731{\ts}/{\ts}H$\alpha$ -- for three
of the eight HzPRGs are consistent with the presence of a
Seyfert{\ts}2 nucleus; the [O
III]{\ts}$\lambda$5007{\ts}/{\ts}H$\beta$ and [S II]{\ts}$\lambda
\lambda$6716, 6731{\ts}/{\ts}H$\alpha$ ratios and/or limits of the
remaining five galaxies are inconclusive.  Furthermore, all six of
the galaxies for which both $H$- and $K$-band spectra were obtained
have observed [O III]{\ts}$\lambda$5007{\ts}/{\ts}(H$\alpha$+[N II]
$\lambda \lambda$6548, 6583) ratios consistent with Seyfert{\ts}2
ionization.  Much of the inability to detect the weaker emission
lines of [S II]{\ts}$\lambda \lambda$6716, 6731 in three of the
galaxies and H$\beta$ in any of the galaxies may be due to moderate
amounts of dust:  for the two sources with previously measured
Ly$\alpha$ fluxes, the observed Ly$\alpha${\ts}/{\ts}H$\alpha$ ratios
are $\sim${\ts}1.5, much less than the value of 16 expected for gas
in a dust-free medium photoionized by a hard, nonthermal continuum.
If such a discrepancy is due solely to dust, this ratio translates
into $A_V \sim${\ts}0.5--1.0 mag (depending on which extinction curve
-- Milky Way, SMC, LMC -- is used) at the rest-frame optical
wavelengths of the galaxies, and a corresponding factor of
$\sim${\ts}1.6--2.5 reduction in optical flux.

None of the eight HzPRGs at $z = 2.2-2.6$ HzPRGs have broad ($\Delta
v_{\rm FWHM} > 1500$ km s$^{-1}$) emission-line cores, and it is not
clear whether any have broad emission-line wings.  However, the
near-infrared spectrum of 3C{\ts}22, a $z = 0.937$ radio galaxy with
1{\ts}$\mu$m luminosity comparable to that of the radio galaxies at
$z=2.2-2.6$ and a radio luminosity only 3--5 times less, shows direct
evidence for broad H$\alpha$ emission wings.  Such a feature is
indicative of the presence of a partially obscured Seyfert{\ts}1
nucleus.  Given that 3C{\ts}22 is at $\sim${\ts}1/3 the luminosity
distance of the sample of HzPRGs at $z = 2.2-2.6$, a thorough search
for such a faint feature in the more distant galaxies may require
8--10 meter-class telescopes.

These new data, along with recent UV-to-optical polarimetry showing
evidence of high polarization in many HzPRGs, provide evidence that
many HzPRGs are predominantly ionized by an active nucleus, and that
a significant fraction of their SED may be due to non-thermal
emission from an active galactic nucleus.

\end{abstract}

\keywords{galaxies: active---galaxies: ISM}

\section{Introduction}

High redshift powerful radio galaxies (HzPRGs:  $P_{\rm 408 MHz} >
10^{27} h^{-2}${\ts}W Hz$^{-1}$, $z
\gtrsim${\ts}1)\footnote{Throughout this paper, $H_o = 100 h${\ts}km
s$^{-1}$ Mpc$^{-1}$ and $q_o = 0.5$ are assumed.} are potentially an
important probe of galaxy evolution at early epochs.  The recent
detections of strong emission lines from HzPRGs at observed
near-infrared wavelengths (e.g., Rawlings, Eales, \& Lacy 1991;
McCarthy, Elston, \& Eisenhardt 1992; Eales \& Rawlings 1993, 1996)
has provided a new diagnostic tool with which to study these objects,
although the nature of line emission in HzPRGs is currently a subject
of great debate.  Part of the difficulty in interpreting line
emission in HzPRGs has been that the use of standard optical
diagnostic emission-line ratios (e.g. Baldwin, Phillips, \& Terlevich
1981; Veilleux \& Osterbrock 1987), commonly used to differentiate
between thermal ionization and low or high-excitation ionization from
a non-thermal source, has until recently been problematic because
rest-frame optical emission from galaxies at $z >${\ts}1 is
redshifted to near-infrared wavelengths at current epochs.

A new generation of infrared spectrographs has now made it possible
to begin systematic studies of the dominant ionization mechanisms in
HzPRGs via their rest-frame optical emission-line spectra.  In
particular, the new {\it K}-band spectrograph (KSPEC: Hodapp et al.
1994) on the University of Hawaii (UH) 2.2{\ts}m telescope on Mauna
Kea provides the unique capability of simultaneous coverage of the
$\sim${\ts}1.0--2.4{\ts}\micron$~~$ wavelength band at typical
spectral resolution $\lambda${\ts}/{\ts}$\Delta \lambda$
{\ts}$\sim${\ts}700, while the infrared spectrograph (CGS4) on the
3.8{\ts}m United Kingdom Infrared Telescope (UKIRT) on Mauna Kea can
provide coverage of either the full {\it J}, {\it H}, or {\it K}-band
near-infrared windows at a spectral resolution of
$\lambda${\ts}/{\ts}$ \Delta \lambda${\ts}$\sim${\ts}860.

A nearly complete sample of HzPRGs in the redshift range $z=2.2-2.6$
has been selected. An attempt has been made, insomuch as possible, to
obtain near-infrared spectra with similar sensitivity and coverage in
rest-frame wavelength. The choice of the redshift range and
wavelength coverage was designed to maximize the number of rest-frame
optical diagnostic lines observed.  As far as the author is aware,
this is the first such study that obtains both {\it H}- and {\it
K}-band spectra for a fairly complete sample of HzPRGs.

The outline of this paper is as follows: Sample selection and
observing procedures are discussed in \S{\ts}2 and \S{\ts}3
respectively.  Data reduction methods are summarized in \S{\ts}4. The
discussion in \S{\ts}5 focuses on the observed emission-line spectra
and calculated extinction in the HzPRGs, as well as on how the
properties of HzPRGs compare with the properties of local radio
galaxies, and optically selected Seyferts, LINERs, and H{\ts}II
region-like galaxies.

\section{Sample Selection}

The study of HzPRGs at near-infrared wavelengths from the ground is
optimized for redshifts in the range $z \sim 2.2-2.6$ which
simultaneously places the largest number of rest-frame optical
diagnostic emission lines (e.g. H$\beta$,
[O{\ts}III]{\ts}$\lambda$5007, H$\alpha$,
[N{\ts}II]{\ts}$\lambda\lambda$6548, 6583, and
[S{\ts}II]{\ts}$\lambda \lambda$6716, 6731 - hereafter [S
II]{\ts}$\lambda$6724) directly in the {\it H}-band and {\it K}-band
atmospheric windows.  When this project was started in early 1993,
relatively few HzPRGs were in the literature with redshifts in this
range and with $\delta > 0^\circ$ (as required by this study for ease
in observing from Mauna Kea). Nine galaxies were ultimately selected
- five of these objects (B3{\ts}0731+438, 3C{\ts}257, 53W002,
MG{\ts}1744+18, and 4C{\ts}40.36) were selected from a list compiled
by Eales \& Rawlings (1993), who presented observed {\it K}-band
spectra of HzPRGs with relatively low sensitivity, and four
(TX{\ts}0200+015, TX{\ts}0828+193, 4C{\ts}48.48, and 4C{\ts}23.56)
were selected from the ultra-steep spectrum (USS) radio galaxy survey
(e.g., R\"{o}ttgering 1993; van Ojik 1995; R\"{o}ttgering et al.
1995; Chambers et al. 1996).  With the exception of 53W002, all of
these HzPRGs were observed as part of this survey.  The positions and
redshifts of the eight observed sources are listed in Table 1.

The final observing list included one additional object, 3C{\ts}22,
which was added during the very first phases of this study.  The
galaxy 3C{\ts}22 has a 1{\ts}$\mu$m luminosity comparable to the
other eight HzPRGs in the sample, and a radio luminosity only 3--5
times fainter, however it has a redshift ($z \sim 0.94$) only
approximately half that of the others.  It was originally included in
the list mainly as a hedge against the possibility that most of the
higher redshift objects would be too faint for obtaining spectra with
the signal-to-noise ratio necessary to make reasonable measurements
of emission-line ratios.  If this had indeed turned out to be the
case, a more limited spectroscopic study of HzPRGs at redshifts $z
\sim 1$ would have been carried out.  However, both KSPEC and CGS4
provided sufficient sensitivity for studying the $z = 2.2-2.6$
sources, although the data obtained for 3C{\ts}22 did have a
significantly higher signal-to-noise ratio than all of the other
sources.  The data for 3C{\ts}22 is therefore interesting in its own
right as an example of how the data for the higher redshift sources
might appear if higher sensitivity near-infrared data is eventually
obtained.

\section{Observations}

In planning the observations of the sample of HzPRGs, the first
priority was to obtain simultaneous {\it H}-band and {\it K}-band
spectra with KSPEC, and then to obtain {\it H}-band and {\it K}-band
spectra using CGS4 of those sources that, for whatever reason, could
not successfully be observed with KSPEC.  In the end, due to weather
and equipment availability, three sources were observed with both
spectrometers and five with CGS4 alone. The galaxy 3C{\ts}22 was
observed only with KSPEC.

Table 1 presents a complete journal of the near-infrared
observations.  {\it K}-band spectra were obtained of eight HzPRGs
with better sensitivity than any previously published {\it K}-band
data, {\it H}-band spectra were obtained for five sources, and one
object was also observed at {\it J}-band.  Except for the recent
publication of {\it H}-band data for one source, these are the first
{\it H}-band and {\it J}-band spectra of HzPRGs in this redshift
range published to date.  Observing parameters are also listed in
Table 1.  Details of the observing procedures at each of the two
telescopes used are discussed separately below.

\subsection{UH 2.2m Telescope}

Observations with the infrared spectrograph, KSPEC (Hodapp et al.
1994), on the UH 2.2m Telescope were made during four observing
periods between 1993 July and 1994 December.  KSPEC is a
cross-dispersed echelle spectrograph configured to cover the
wavelength range 1.1--2.5{\ts}$\mu$m in three orders ({\it J, H,} and
{\it K}) on a 256$\times$256 NICMOS-3 HgCdTe detector array.  The
wavelength resolutions (at 2.2 $\mu$m) for each of the observations
are listed in Table 1.  Additionally, the wavelength range
0.7--1.0{\ts}$\mu$m is also dispersed in several orders on the array.
However, the order crowding at 0.7--1.0{\ts}$\mu$m is such that only
spectra of point-like sources can effectively be extracted, and this
wavelength range is also compromised due to the fact that KSPEC is
only in focus at $\lambda >${\ts}0.85{\ts}$\mu$m.  In addition to its
spectral capabilities, KSPEC also provides simultaneous imaging of
approximately one square arcminute of sky around the slit on a second
NICMOS-3 array.  For the 1994 December observing period, the new UH
tip-tilt system (Jim et al. 1997) was also implemented, notably
improving the tracking, offsetting, and seeing of the observations
and data.

For the 1995 November observing period, an upgraded version of KSPEC
was used in combination with the UH tip-tilt system.  The new device
uses a HAWAII 1024$\times$1024 HgCdTe array as the spectral detector,
and provides coverage over the wavelength range 0.81--2.54{\ts}$\mu$m
without confusion due to overlapping orders.

Each observing night began with a series of dome flats with
incandescent lights turned on, then off. Once the telescope was
guiding with the source in the slit, a 180 sec exposure was taken,
followed by a second 180 sec exposure with the slit positioned
10\arcsec~ off-source. For all observations, except during 1995
November, each spectrum exposure was accompanied by a 140 sec image
exposure. No image exposures were taken for the 1995 November
observations except to occasionally check the position of stars in
the field. For observations prior to 1994 December, the observing
pattern was source-sky-source-sky ..., but the pattern was changed to
source-sky-sky-source ... for the 1994 December observations to
minimize spurious features created by changing sky conditions.
Observations of standard AV stars near each source were made for flux
calibration and to remove telluric lines.  Observations of an argon
lamp were used for wavelength calibration.

\subsection{United Kingdom 3.8m Infrared Telescope}

Observations were made with the infrared spectrometer CGS4 on UKIRT
during the observing periods 1993 August, 5--9 and 1994, October 3--7
.  CGS4 is a 1--5{\ts}$\mu$m 2D grating spectrometer with a
58$\times62$ InSb array.  For all observations, a 75 line mm$^{-1}$
grating was used in first order.  The wavelength resolutions and
sampling scheme used for each source is listed in Table 1.  For most
sources, sampling was done over 2 pixels in 6 or 8 steps in the
wavelength direction.  A one pixel-wide (1.5\arcsec) slit was used
for all observations.

Observations with an upgraded version of CGS4 were made during four
observing periods from 1995, May to 1996, April.  The 1995, May
observations were done under shared-risk time.  The new detector was
a 256$\times$256 InSb array.  A 75 line mm$^{-1}$ grating in first
order was used for the first three observing periods, and a 150 line
mm$^{-1}$ grating in first order was used for the last observing
period.  A one pixel-wide (1.2\arcsec) slit was used for all
observations except for the B3{\ts}0731+438 and 3C{\ts}257
observations which were obtained using a two pixel-wide 2.5\arcsec~
slit.

Spectral data were taken using exposure times between 60 to 540 sec,
with 20 to 90 sec of each exposure made at each detector position
during the sampling.  The slit was then moved 15\arcsec--30\arcsec~
in the spatial direction on the array for the next multi-sampling
exposure. Thus, the source appears on the array as a positive and
negative spectrum after the first frame is subtracted from the
second.  The source-sky observation sequence was the same as that
used in the 1994, December KSPEC observations. Flux calibrations were
again made using AV standard stars located near each source, and a
krypton lamp was used for wavelength calibration.

For the 1995 May observations of MG{\ts}1744+18, a problem was
encountered; the source flux of the positive spectrum was more than
that of the negative spectrum flux, implying that the source was only
partially in the slit in the second position.  Alternatively, the
flux of the positive A star spectrum was less than the negative
spectrum flux. The slit also appeared to be tilted relative to the
array as evident by the position of the OH sky lines.  Because of
this, only half the data had enough flux to be usable, and no
photometry was possible for MG{\ts}1744+18.

\section{Results}

All data reduction was done within IRAF.  For KSPEC data the
procedure was as follows:  Sky frames were first subtracted from the
source frames.  The result was then divided by an appropriate flat
field and then averaged.  Since there were only a few bad pixels, and
none were on the spectral area containing the source spectrum, the
bad pixels were individually set to zero before the spectral orders
were extracted with the APALL package. The extracted spectra were
then wavelength calibrated and divided by the standard star spectrum
(which had been averaged, flatfielded, and extracted in the same
manner as the source spectrum) to remove any instrumental effects and
atmospheric lines.  Because of the narrow width of the slit, no
photometry was determined for any of the KSPEC observations. Finally
the spectrum was multiplied by a Planck blackbody spectrum with the
same temperature as the standard star. The CGS4 data were reduced in
a same manner as the KSPEC data, except that flat fielding and
masking of bad pixels was done automatically after each observation.
Photometry was determined for all galaxies except B3{\ts}0731+438,
which was observed at the end of a observing period of variable
weather, and MG{\ts}1744+18 (see \S 3.2).

The near-infrared spectra for all eight HzPRGs at $z = 2.2-2.6$ that
were observed are plotted in Figure 1. Measured emission-line
properties are summarized in Table 2. For most of the sources, [O
III]{\ts}$\lambda\lambda$4959, 5007, H$\alpha$+[N II]{\ts}$\lambda
\lambda$6548, 6583, and [S II]{\ts}$\lambda$6724 were detected. The
emission line [O II]{\ts}$\lambda$3727 was also detected in
4C{\ts}48.48.  H$\beta$ and [O I]{\ts}$\lambda$6300 were not detected
in any of the HzPRGs.  Taking into account the signal-to-noise ratio
of any individual galaxy spectrum, all of the HzPRG emission-line
spectra plotted in Figure 1 (with the exception of the anomalous
source TX{\ts}0828+193SW) look similar, and the lines have similar
widths.

For 4C{\ts}23.56 there is a discrepancy in the H$\alpha$ flux
measured from two separate observing periods. During the 1993, August
observations, the telescope was positioned on the source by
offsetting from a nearby, faint star.  For the 1994, October
observations, the telescope was first peaked-up on a relatively
nearby star whose position is known to high accuracy, then the
telescope was slewed to the radio coordinates of the source.  Because
the H$\alpha$ flux measured in the latter observations is three times
stronger than the line measured in the former, the position
uncertainty of the faint offset star is most likely a large fraction
of the slit width and the 1993, August data will be ignored in
further discussion of this source.

For 4C{\ts}40.36 the observing conditions were variable for the
period the data were obtained.  Photometry for the standard star
(HD{\ts}166208) observed during the night in question was compared to
the standards (BS{\ts}7503 and BS{\ts}8143) observed on the two
photometric nights, and the flux of HD{\ts}166208 was found to be
20{\ts}\% lower than the flux expected under photometric conditions.
The spectrum in Figure 1 and the emission-line flux values in Table 2
for 4C{\ts}40.36 have been scaled appropriately to take this factor
into account.

Figure 2 shows the spectrum of the lower redshift radio galaxy
3C{\ts}22 that was obtained during 1994, December.  The
signal-to-noise ratio of this spectrum is significantly better than
that of the objects in Figure 1.  The H$\alpha$+[N II]{\ts}$\lambda
\lambda$6548, 6583 line wings for 3C{\ts}22 are remarkably broad,
characteristic of Seyfert{\ts}1 galaxies.  Emission-line ratios for
3C{\ts}22 are also summarized in Table 2, but their discussion will
be postponed until after the discussion of the higher redshift
objects.

Several of these sources have been observed by others (McCarthy et
al. 1992; Eales \& Rawlings 1993, 1996).  Of the two HzPRGs for which
photometric values can be compared with previous measurements, the
H$\alpha$+[N II] $\lambda \lambda$ 6548, 6583 flux value of
$2.9\times10^{-18}$ W m$^{-2}$ reported by Eales \& Rawlings (1996)
for 3C{\ts}257 agrees well with the value of $2.6\times10^{-18}$ W
m$^{-2}$ in Table 2, while the value of $3.4\times10^{-18}$ W
m$^{-18}$ for 4C{\ts}40.36 (Table 2) is higher than value of
$2.6\times10^{-18}$ W m$^{-2}$ reported by Eales \& Rawlings (1993).
The Eales \& Rawlings (1993, 1996) observations made use of a
$3\arcsec \times 3\arcsec$ aperture, thus the agreement in 3C{\ts}257
line flux and lack of agreement in 4C{\ts}40.36 line flux may
indicate that the H$\alpha$+[N II] $\lambda \lambda$ 6548, 6583
emission-line region of 3C{\ts}257 is compact and that of
4C{\ts}40.36 is extended, or that the H$\alpha$+[N II] $\lambda
\lambda$ 6548, 6583 flux for 4C{\ts}40.36 presented in this paper is
an overestimate.  The resolution, sampling, and integration times
(Table 1) for the spectra presented in this paper have allowed for a
better determination of rest-frame optical line profiles of $z>2$
HzPRGs than previously published.  In addition, the H$\alpha$+[N II]
$\lambda \lambda$ 6548, 6583 profile of 3C{\ts}22 (Figure 2) is
consistent with recently published data of this source (Economou et
al.  1995; Rawlings et al. 1995).

\section{Discussion}

\subsection{Emission-Line Diagnostics}

Figure 3 shows the log ([O III]{\ts}$\lambda$5007{\ts}/{\ts}H$\beta$)
vs.  log ([S II]{\ts}$\lambda$6724{\ts}/{\ts}H$\alpha$) diagram
commonly used to distinguish between galaxies with Seyfert, LINER,
and H{\ts}II-region emission-line spectra.  Emission from [S
II]{\ts}$\lambda$6724 can emanate from ionized hydrogen regions, as
well as from semi-ionized regions where collisional ionization is
significant.  Enhancement of forbidden lines such as [O
I]{\ts}$\lambda$6300, [N II]{\ts}$\lambda \lambda$6548, 6583, and [S
II]{\ts}$\lambda$6724 occurs in AGNs because, unlike H{\ts}II
regions, they have extended partially-ionized zones created by an
excess of X-rays (the absorption cross sections of neutral hydrogen,
helium and all ions are small for X-rays, thus X-rays tend to escape
the ionized region before interacting: Veilleux \& Osterbrock 1987).
[O III]{\ts}$\lambda$5007 is a high ionization line photoionized by
UV photons, and thus tends to be strong in Seyfert galaxies.

Instead of attempting to determine the relative contributions of
H$\alpha$ and the [N II]{\ts}$\lambda \lambda$6548, 6583 doublet to
the blended H$\alpha$+[N II]{\ts}$\lambda \lambda$6548, 6583 complex,
the smallest ratio of [N II]{\ts}$\lambda$6583{\ts}/{\ts}H$\alpha$ (=
0.19) observed for a sample of luminous infrared galaxies (Kim et al.
1995) has been adopted and plotted (large circles with embedded
numerals) for all of the HzPRGs in the sample\footnote{ It is
unlikely, given the widths of the [O III]{\ts}$\lambda$5007 lines,
that the H$\alpha$+[N II]{\ts}$\lambda \lambda$6548, 6583 complexes
are pure H$\alpha$.  However, if the complexes were purely H$\alpha$,
their observed breadth ($\Delta v_{\rm FWHM} \gtrsim$ 1000 km
s$^{-1}$) would be evidence against H{\ts}II region-like emission.}.
Choosing such a ratio tends to bias the data towards the
H{\ts}II-region portion of the diagram.  To illustrate the extent to
which the data are affected by this ratio, the values that result
from assuming [N II]{\ts}$\lambda$6583{\ts}/{\ts}H$\alpha$ = 1.0, the
value adopted by Eales \& Rawlings (1993) based on the predictions of
photoionization models and the observed value for low-redshift
powerful radio galaxies (LzPRGs), and [N
II]{\ts}$\lambda$6583{\ts}/{\ts}H$\alpha$ = 3.74, the largest value
observed by Kim et al. (1995), have also been plotted as small,
filled-in circles.  Note that in all of the $H$-band spectra except
for the spectrum of 4C{\ts}48.48, the estimated H$\beta$ upper limits
(3{\ts}$\sigma$) provide weak constraints on the lower limit of [O
III]{\ts}$\lambda$5007{\ts}/{\ts}H$\beta$.  Thus, in all sources
except 4C{\ts}48.48 and 4C{\ts}40.36, the [O
III]{\ts}$\lambda$5007{\ts}/{\ts}H$\beta$ ratio is determined
assuming H$\alpha${\ts}/{\ts}H$\beta$ $\ge${\ts}3, and thus [O
III]{\ts}$\lambda$5007{\ts}/{\ts}H$\beta$
$\ge${\ts}3{\ts}$\times${\ts}[O
III]{\ts}$\lambda$5007{\ts}/{\ts}H$\alpha$ (the upper limit for [O
III]{\ts}$\lambda$5007{\ts}/{\ts}H$\beta$ in 4C{\ts}40.36 is taken
from Iwamuro et al. 1996).  Three of the eight HzPRGs plotted in
Figure 3 clearly fall in the Seyfert region of the plot.  The [O
III]{\ts}$\lambda$5007{\ts}/{\ts}H$\beta$ and [S
II]{\ts}$\lambda$6724{\ts}/{\ts}H$\alpha$ ratios and/or limits for
TX{\ts}0200+015, TX{\ts}0828+193, 3C{\ts}257, MG{\ts}1744+18, and
4C{\ts}23.56 are inconclusive.

Additional information about the ionization mechanism is obtained by
examining the [O III]{\ts}$\lambda$5007{\ts}/{\ts}(H$\alpha$+[N II]
$\lambda \lambda$6548, 6583) ratios.  Table 3 lists the average value
for this ratio as a function of emission-line classification for
galaxies in the {\it IRAS} Bright Galaxy Sample (BGS) and a sample of
``warm'' {\it IRAS} galaxies (Kim et al. 1995; Veilleux et al. 1995),
as well as for a sample of LzPRGs with Seyfert{\ts}2 emission-line
spectra (i.e., Cygnus A: Osterbrock \& Miller 1975; PKS{\ts}1345+12:
Grandi 1977; 3C98, 3C192, 3C327: Costero \& Osterbrock 1977).
Despite the obvious scatter due presumably to extinction and possible
metallicity effects, the ratio provides a notable separation between
Seyfert galaxies and those of the H{\ts}II and LINER class.  A
comparison of these low-redshift, active galaxies with the six HzPRGs
at $z = 2.2-2.6$ for which both $H$- and $K$-band spectra have been
obtained show all six HzPRGs to have [O
III]{\ts}$\lambda$5007{\ts}/{\ts}(H$\alpha$+[N II] $\lambda
\lambda$6548, 6583) ratios consistent with Seyfert{\ts}2
galaxies\footnote{The [O III]{\ts}$\lambda$5007 measurement of
4C{\ts}40.36 by Iwamuro et al. (1996), obtained with a $PA = 0$ and a
1.5\arcsec$\times$60\arcsec$\,$ aperture, is included because their
$H$-band image shows most of the flux to be within 1.5\arcsec$\,$ of
the center of the galaxy. The [O III]{\ts}$\lambda$5007 flux is
$3.86\times10^{-18}$ W m$^{-2}$, and thus the observed [O
III]{\ts}$\lambda$5007{\ts}/{\ts}H$\alpha$+[N II] $\lambda
\lambda$6548, 6583 ratio for 4C{\ts}40.36 is 1.1.}.  The average
value of the [O III]{\ts}$\lambda$5007{\ts}/{\ts}(H$\alpha$+[N II]
$\lambda \lambda$6548, 6583) ratio for these six HzPRGs is also
listed in Table 3.  Thus, the dominant source of ionization for these
galaxies appears to be no different from that observed in most
low-redshift, narrow-line radio galaxies.

Because the H$\alpha$ and [N II]{\ts}$\lambda \lambda$6548, 6583
lines are blended, a direct comparison of the low
ionization-to-H$\alpha$ line ratio of these high-redshift galaxies to
their low-redshift counterparts cannot be made. An alternative is to
compare the [S II]{\ts}$\lambda$6724{\ts}/{\ts}(H$\alpha$+[N II]
$\lambda \lambda$6548, 6583) ratio.  For the 5 HzPRGs with [S
II]{\ts}$\lambda$6724 detections, [S
II]{\ts}$\lambda$6724{\ts}/{\ts}(H$\alpha$+[N II] $\lambda
\lambda$6548, 6583) {\ts}$= 0.25 \pm 0.07$, similar to the mean value
of $0.27 \pm 0.05$ observed for a sample of LzPRGs with Seyfert{\ts}2
emission-line spectra.  Changes in properties such as the elemental
abundances of the host galaxy, the shape of the ionizing spectrum,
and the electron density of the gas being ionized will act to vary
this ratio (note that extinction has little effect on this ratio
because these emission lines have similar wavelengths).  Table 3
contains a summary of average [S
II]{\ts}$\lambda$6724{\ts}/{\ts}(H$\alpha$+[N II] $\lambda
\lambda$6548, 6583), H$\alpha${\ts}/{\ts}[S II]{\ts}$\lambda$6724,
H$\alpha${\ts}/{\ts}[N II] $\lambda \lambda$6548, 6583, and [N
II]{\ts}$\lambda \lambda$6548, 6583{\ts}/{\ts}[S
II]{\ts}$\lambda$6724 ratios as a function of emission-line
classification.  Essentially, [S
II]{\ts}$\lambda$6724{\ts}/{\ts}(H$\alpha$+[N II] $\lambda
\lambda$6548, 6583) is a function of H$\alpha${\ts}/{\ts}[S
II]{\ts}$\lambda$6724 and [N II]{\ts}$\lambda \lambda$6548,
6583{\ts}/{\ts}[S II]{\ts}$\lambda$6724, where H$\alpha${\ts}/{\ts}[S
II]{\ts}$\lambda$6724 appears to account for most of the variation in
the average.  On average, [S
II]{\ts}$\lambda$6724{\ts}/{\ts}(H$\alpha$+[N II] $\lambda
\lambda$6548, 6583) is larger for LINERs and Seyfert 2 galaxies than
in H{\ts}II region-like galaxies (see Table 3), in part because the
former have more extended semi-ionized regions (i.e.,
H$\alpha${\ts}/{\ts}[S II]{\ts}$\lambda$6724 decreases as the extent
of the semi-ionized region increases).  Variations in the metal
abundance in the galaxies in the sample undoubtedly contribute to the
scatter.  For example, a substantial decrease in the metal abundance
would cause a substantial decrease in [S
II]{\ts}$\lambda$6724{\ts}/{\ts}(H$\alpha$+[N II] $\lambda
\lambda$6548, 6583) (i.e., H$\alpha${\ts}/{\ts}[S
II]{\ts}$\lambda$6724 increases as the metal abundance decreases),
and vice versa.  The observation that LzPRGs and HzPRGs have similar
[S II]{\ts}$\lambda$6724{\ts}/{\ts}(H$\alpha$+ [N II]{\ts}$\lambda
\lambda$6548, 6583) ratios may indicate that, even though HzPRGs are
at substantially higher lookback times and possess higher radio
luminosities, the properties of the ionizing source and the ionized
gas are similar.

It is worthwhile to consider whether the gas in the HzPRGs could be
ionized by supernova remnants (SNRs). Indeed, there are a few SNRs
that exhibit emission-line ratios similar to Seyfert{\ts}2 galaxies.
A conservative estimate of the SNR rate required to explain the
observed emission lines can be determined by considering the extreme
SNR in NGC{\ts}6946, which has log([O
III]{\ts}$\lambda$5007{\ts}/{\ts}H$\beta$) = 0.85 and log([S
II]{\ts}$\lambda$6724{\ts}/{\ts}H$\alpha$) $= -0.06$ (Blair \& Fesen
1994).  The flux in the [O III]{\ts}$\lambda$5007 line is
$2.5\times10^{-17}$ Watts m$^{-2}$.  At a distance corresponding to
$z = 2.4$, the SNR would have a flux of $\sim 2.5\times10^{-17}$(5.1
Mpc / $9330h^{-1}$ Mpc)$^2$ = $7.5\times10^{-24}h^2$ Watts m$^{-2}$.
Thus, to produce the average [O III]{\ts}$\lambda$5007 flux (i.e.
$3\times10^{-18}$ Watts m$^{-2}$) observed in the HzPRGs in the
sample would require $4\times10^5h^{-2}$ SNRs.  Adopting a
conservative estimate of 5000{\ts}yr for the SNR lifetime, HzPRGs
would have to produce SNRs at an implausible rate of
$80h^{-2}${\ts}yr$^{-1}$ to maintain their [O III]{\ts}$\lambda$5007
flux.

\subsection{$L_{\rm [O III]{\ts}\lambda \lambda4959, 5007}$ vs. $P_{\rm 151 MHz}$}

Previous authors have used the [O III]{\ts}$\lambda \lambda$4959,
5007 luminosity versus the observed 151 MHz radio power for HzPRGs in
an attempt to infer a causal relationship between the ionization
source for the gas and the source of the radio emission.  Rawlings et
al. (1989) have compiled data primarily for LzPRGs, and more
recently, Eales \& Rawlings (1993, 1996) have added data for HzPRGs.
Figure 4a is a plot adapted from Eales \& Rawlings (1996) of the [O
III]{\ts}$\lambda \lambda$4959, 5007 luminosity versus the observed
151 MHz radio power for an unbiased 3C sample of FR{\ts}II radio
galaxies with $z <${\ts}0.5 and a collection of HzPRGs.  Data for the
HzPRGs TX{\ts}0200+015, B3{\ts}0731+438, TX{\ts}0828+193,
4C{\ts}48.48, and 4C{\ts}23.56 have also been plotted.

A simple interpretation of Figure 4a would be that there appears to
be a correlation between the [O III] $\lambda \lambda$ 4959, 5007
emission-line luminosity and the radio power (such a correlation
would appear to be even tighter perhaps for the HzPRGs than for the
LzPRGs), and that a common excitation source (e.g. the central AGN)
is responsible for both.  However, the real answer is clearly not so
simple.  The apparent strong correlation between the [O III] $\lambda
\lambda$ 4959, 5007  emission-line luminosity and the radio power,
$P_{\rm 151 MHz}$, in Figure 4a is almost entirely an artifact of
distance, as evidence by the plot of the ratio $L_{\rm [O
III]{\ts}\lambda \lambda4959, 5007} / P_{\rm 151 MHz}$ vs.  $P_{\rm
151 MHz}$ (Figure 4b) which shows no correlation in the sample as a
whole.  However, note the apparent strong correlation exhibited in
Figure 4b by the HzPRGs (i.e., all sources with $P_{\rm 151 MHz} >
10^{27}$ W Hz$^{-1}$ sr$^{-1}$ and $L_{\rm [O III]{\ts}\lambda
\lambda4959, 5007} / P_{\rm 151 MHz} > 10^8$).  Indeed, for the most
powerful radio galaxies at $z >${\ts}0.5 the morphology of the
emission-line gas is aligned with the radio axis (McCarthy et al.
1987), which suggests that a correlation between the two quantities
exists.

The extraordinary strength of the emission lines in HzPRGs is also
evident from the data in Table 4: of the TX and 4C galaxies in the
sample for which high signal-to-noise ratio $H$- or $K$-band
magnitudes were made available from Keck imaging (L. Armus, private
communication), the emission lines appear to contribute anywhere
between 25--98{\ts}\% of the broad-band near-infrared light.  These
percentages are substantially higher than those observed in the most
powerful radio galaxies in the local Universe, but are consistent
with other HzPRGs with published near-infrared spectra (McCarthy et
al. 1992; Eales \& Rawlings 1993, 1996).  Such high emission-line
luminosities illustrate the danger of interpreting the magnitudes and
morphologies of high radio-power galaxies as being purely stellar in
origin (see Eales \& Rawlings 1993, 1996 for discussions on this
issue ).  Evidence for possible AGN contamination of the broad-band
morphologies of high radio-power galaxies has been demonstrated by
Dunlop \& Peacock (1993), who show that the rest-frame 1.1 $\mu$m
flux of a sample of 3CR galaxies tend to be more extended and aligned
with the radio axis than a sample of lower radio-power Parkes
galaxies in the same redshift range ($z \sim 1$).  Such extended and
aligned morphologies are more pronounced at rest-frame UV wavelengths
and are undoubtedly connected with the energetics of the central
engine, and/or due to an optically-thick, circumnuclear dust/gas
torus extinguishing light perpendicular to the radio axis.
Rest-frame UV spectropolarimetry (Dey \& Spinrad 1995; Cimatti et al.
1996; Manzini \& di Serego Alighieri 1996 and references therein) and
optical imaging polarimetry (Knopp \& Chambers 1997; Knopp 1997),
show further evidence of AGN contamination of broad-band light in
many HzPRGs.\footnote{There is one example of an HzPRG, 4C 41.17 at
$z=3.8$, that does not appear to have polarized UV light (Dey et al.
1997).} Indeed, four of the galaxies discussed in this paper
(TX{\ts}0200+015, B3{\ts}0731+438, TX{\ts}0828+193, and 4C{\ts}23.56)
show rest-frame optical polarizations of 10--45\% (Knopp \& Chambers
1997; Knopp 1997), indicating that a significant fraction of the
rest-frame UV-to-optical continuum emission from many HzPRGs may be
scattered/reprocessed AGN light.

\subsection{TX{\ts}0828+193SW: A Component Dominated by Continuum-Emission}

In some HzPRGs, there appear to be ``components'' that are genuinely
dominated by continuum emission at observed optical and near-infrared
wavelengths.  Figure 1 shows two extractions of the southwestern
``component'' of TX{\ts}0828+193. Unlike the northeastern component,
TX{\ts}0828+193SW has no notable strong emission lines, but has
comparatively strong continuum emission.  Spectroscopy of the radio
galaxy at wavelengths near 4000 \AA$~$ (observer-frame) reveals
strong Ly$\alpha$, C{\ts}IV, He{\ts}II, and C{\ts}III] emission lines
in the NE component, but no emission lines in the SW component (van
Ojik 1995; R\"{o}ttgering et al. 1997).  Radio, $R$-band imaging
data, and near-infrared imaging data taken of TX{\ts}0828+193 show
the radio core to be coincident with the NE component and the SW
component to lie $\sim$ 5\arcsec~ away from the radio core, but still
along the radio axis of the galaxy (R\"{o}ttgering et al. 1995; Knopp
\& Chambers 1997; see Figure 5 in the Appendix).  This raises the
possibility that either TX{\ts}0828+193SW is a cloud being
illuminated by continuum emission emanating from TX{\ts}0828+193NE,
or that a buried AGN residing in TX{\ts}0828+193SW is ionizing gas in
TX{\ts}0828+193NE.  The two best examples of ionization of
off-nuclear knots in low-redshift radio galaxies are Coma A (van
Breugel et al. 1985) and PKS 2152-69 (Tadhunter et al. 1987).  In the
case of Coma A, the nuclear region has strong continuum emission and
relatively weak lines, whereas the off-nuclear knot has relatively
strong line emission. In the case of PKS 2152-69, both the nucleus
and the off-nuclear knot have notable continuum emission, but the
line emission from the off-nuclear knot, especially [O III]$\lambda$
5007, is stronger.  Such data would imply that the AGN is in
TX{\ts}0828+193SW and is ionizing the northeastern component.  Recent
multi-wavelength, broad-band polarimetry measurements of
TX{\ts}0828+193 show evidence for TX{\ts}0828+193NE having
polarization consistent with scattering due to dust (Knopp \&
Chambers 1997).

Because TX{\ts}0828+193SW has no emission-line or absorption features
from which its redshift can be determined, there exists the
possibility that it is simply an unrelated object along the line of
sight.  Though the chance of such a superposition is low, there is
precedent for concern, most notably the foreground star coincident
with 3C368 (Hammer, Le F\`{e}vre, \& Proust 1991).  In the recently
discovered HzPRG MG{\ts}1019+0535, a double-component morphology is
observed, one component of which shows no sign of strong line
emission (Dey, Spinrad, \& Dickinson 1995). The authors argue
strongly in favor of the lineless component being a foreground
object, but unlike TX{\ts}0828+193, the two `components' of
MG{\ts}1019+0535 are orthogonal to the radio axis.  Given the
geometry of TX{\ts}0828+193, the polarimetry measurements, and that
such double-component structures aligned with the radio axis are
observed in other radio galaxies (e.g. MRC{\ts}0406-24: Eales \&
Rawlings 1993), TX{\ts}0828+193SW is most likely at the redshift of
the radio source.

\subsection{Dust}

The evidence to date that HzPRGs as a class contain substantial
amounts of dust is mixed, but compelling - while polarimetry
observations show evidence for scattering by dust in several HzPRGs
(Dey \& Spinrad 1995; Cimatti et al. 1996; Manzini \& di Serego
Alighieri 1996 and references therein; Knopp \& Chambers 1997; Knopp
1997), far-infrared/submillimeter bolometry (Golombek, Miley, \&
Neugebauer 1988; Evans et al. 1996; Dunlop et al. 1994; Chini \&
Kreugel 1994; Ivison 1995; Hughes, Dunlop, \& Rawlings 1997) and CO
spectroscopy (Evans et al. 1996; Downes et al. 1996; van Ojik et al.
1997a; Scoville et al. 1997) surveys of HzPRG show evidence for
substantial amounts of dust in only a few sources.  Regardless of
whether or not HzPRGs have comparable or more dust as that inferred
by the SEDs and molecular gas masses of many low-redshift radio
galaxies (e.g, Golombek, Miley, \& Neugebauer 1988; Mirabel, Sanders
\& Kaz\`{e}s 1989; Knapp \& Patten 1991; Impey \& Gregorini 1993;
Mazzarella et al. 1993; Evans 1996), only moderate amounts are
required to notably decrease the observed flux of the optical
(rest-frame) emission lines and continuum.  Specifically, the
extinction in the HzPRGs may be substantial enough such that the
weaker lines of H$\beta$, [O I]{\ts}$\lambda$6300, and [S
II]{\ts}$\lambda$6724, and possible features such as emission-line
wings (see \S 5.5), fall below the detection threshold of the spectra
in Figure 1.

Estimates of extinction in HzPRGs in the sample were made by
comparing the observed line ratios of hydrogen recombination lines
with their intrinsic ratio (i.e., ratios in a dust-free
environment).  Intrinsic ratios for low-density gas photoionized by a
thermal continuum source are H$\alpha${\ts}/{\ts}H$\beta$ = 2.85 and
Ly$\alpha${\ts}/{\ts}H$\alpha$ = 8.10 (Osterbrock 1989). However,
because the evidence presented here and elsewhere (see McCarthy 1993)
implies that gas in HzPRGs is heated by a hard nonthermal continuum,
enhanced Ly$\alpha$ and H$\alpha$ emission in predominantly neutral
regions heated by X-rays must be taken into account (note that
collisions resulting in the emission of Ly$\alpha$ and H$\alpha$ are
comparatively more frequent than those resulting in H$\beta$).  Thus,
the intrinsic line ratios applicable to HzPRGs are
H$\alpha${\ts}/{\ts}H$\beta$ = 3.1 and Ly$\alpha${\ts}/{\ts}H$\alpha$
= 16 (Ferland \& Osterbrock 1985; Osterbrock 1989).

The H$\beta$ emission line was not detected in any of the HzPRG
spectra shown in Figure 1. Thus, the Ly$\alpha${\ts}/{\ts}H$\alpha$
ratio must be used to calculate the extinction for individual
sources.   This ratio can be determined for four of eight sources; in
the case of B3{\ts}0731+438 ($\sim 2.6$) and 3C{\ts}257 ($\sim 0.1$)
by using data from Eales \& Rawlings (1993)\footnote {McCarthy et al.
(1992) have also computed Ly$\alpha${\ts}/{\ts}H$\alpha$ for
B3{\ts}0731+438, as well as for the HzPRG MRC 0406-24. The
discrepancy between the ratios determined by them and those
determined by Eales \& Rawlings (1993) arise from the fact that the
former assume H$\alpha =$ H$\alpha$+[N II] and the latter assume
H$\alpha = \case{1}{2}$ (H$\alpha$+[N II]).}, and in the case of
TX{\ts}0200+015 ($\sim 1.7$) and TX{\ts}0828+193 ($\sim 1.4$) by
using data from Table 2 and R\"{o}ttgering et al. (1997).  The
average observed Ly$\alpha${\ts}/{\ts}H$\alpha$ value determined for
a larger sample of lower-redshift 3CR radio galaxies is
$\sim${\ts}1.5 (McCarthy 1988), and the observed
Ly$\alpha${\ts}/{\ts}H$\alpha$ ratio determined for two other $z>2$
HzPRGs (Eales \& Rawlings 1993) are $\sim 1.4$ (MRC{\ts}0406-24) and
$\sim 1.4$ (53W002).  All of the observed ratios are much less than
the value of 16 expected for gas in a dust-free medium photoionized
by a hard, nonthermal continuum.

The $E(B-V)$ values determined for the $z>2$ HzPRGs B3 0731+438,
3C{\ts}257, MRC 0406-24, and 53W002 using data from Eales \& Rawlings
(1993) are in the range 0.11--0.76\footnote{This is the range of
$E(B-V)$ for these sources assuming the extinction curves for the
Milky Way, the Large Magellanic Cloud, and the Small Magellanic
Cloud, and assuming an [N II]{\ts}$\lambda$6583{\ts}/{\ts}H$\alpha$
ratio of 1.0 and 0.19. See the rest of the paragraph for a detailed
explanation.}.  Using the H$\alpha$ measurements in Table 2, in
combination with the R\"{o}ttgering et al. (1997) Ly$\alpha$
measurements for TX 0200+015 and TX 0828+193, the $E(B-V)$ of these
two galaxies was determined using the following procedure:  Because
the spectral resolution is sufficient to split the H$\alpha$+[N
II]{\ts}$\lambda$6583 complex, the values of H$\alpha$ have been
determined using the three ratios of [N
II]{\ts}$\lambda$6583{\ts}/{\ts}H$\alpha$ plotted on Figure 3 (i.e.,
0.19, 1.0, 3.74).  The observed Ly$\alpha${\ts}/{\ts}H$\alpha$ line
ratios were first corrected for Galactic extinction, then the
extinction in the HzPRGs was determined assuming the extinction
curves for the Milky Way (Savage \& Mathis 1979), LMC (Nandy et al.
1981), and SMC (Prevot et al. 1984); the corresponding values of
$E(B-V)$ are listed in Table 5.  Given the redshift of the
H$\alpha$+[N II]{\ts}$\lambda \lambda$6548, 6583 complex and shape of
the line emission, it seems quite unlikely that the [N
II]{\ts}$\lambda$6583{\ts}/{\ts}H$\alpha$ ratio is as high as 3.74.
Thus, using the values from Table 5 and excluding values calculated
assuming [N II]{\ts}$\lambda$6583{\ts}/{\ts}H$\alpha$ = 3.74,
feasible values of $E(B-V)$ range from 0.14 to 0.39.  Such $E(B-V)$
values translate into $A_V \sim 0.5-1.0${\ts}mag (depending on which
extinction curve -- Milky Way, SMC, LMC -- is adopted) at the
observed infrared wavelengths (i.e., rest-frame optical wavelengths
of HzPRGs at $z=2.2-2.6$), and a corresponding factor of $\sim${\ts}2
reduction in the emission-line and continuum flux.

Although extinction is commonly determined using the luminous
Ly$\alpha$ and H$\alpha$ emission lines (as done above), there are
three concerns in using Ly$\alpha$ for such a measurement.  These
concerns and their relative applicability to the analysis of HzPRGs
are summarized here.  First, if there is a sufficient density of
atomic hydrogen, the number of Ly$\alpha$ photons along the line of
sight can be greatly diminished by resonant scattering or associated
absorption.  One argument against this is that the measured
Ly$\alpha$ emission-line widths are broad enough ($\gtrsim$ few
hundred km s$^{-1}$) such that most of the photons are in the wings
of the line (McCarthy 1996). This can be understood if there exists a
large velocity gradient across the gas that is being ionized -
Ly$\alpha$ radiation emitted from gas at a given radius from the AGN
passes through the gas farther out which is traveling at slower
velocities because, to this gas, the Ly$\alpha$ photons do not appear
to be resonant photons.  However, recent medium resolution
spectroscopy of 15 HzPRGs (van Ojik et al. 1997b), of which TX
0200+015 and TX 0828+193 are included, show evidence in favor of
Ly$\alpha$ absorption.  If the intrinsic Ly$\alpha$ profiles they fit
to the observations are correct, the Ly$\alpha$ emission in these two
galaxies may be diminished by up to 50\%, making $E(B-V)$ lower by
0.10, 0.07, and 0.04 for models assuming a Milky Way, a LMC, and a
SMC extinction curve, respectively.  Second, if the gas is dense
enough, Ly$\alpha$ photons can be destroyed by collisional
de-excitation of the 2$^2P$ state. This is because the Ly$\alpha$
photons must random walk out of the gas, and thus the lifetime of the
2$^2P$ state is lengthened by the number of steps a Ly$\alpha$ photon
must take to escape the gas. However, the densities required for the
collisional de-excitation timescale to be shorter than the radiative
de-excitation lifetime of the 2$^2P$ state are in excess of 10$^{10}$
cm$^{-3}$, a density only believed to be found in the densest areas
of the broad-line region of AGNs (Osterbrock 1989).  Third, if the
alignment of the rest-frame UV morphology and the radio axis, as
observed in many HzPRGs at $z >${\ts}0.5, is the result of external
illumination of gas and dust clouds by radiation from a central
engine, these clouds most likely reflect the majority of the
Ly$\alpha$ photons (e.g., McCarthy 1996).  Given such a geometry, the
Ly$\alpha$ luminosity would provide little information on the amount
of dust present.  However, such a process would also tend to raise,
not lower, the Ly$\alpha$/H$\alpha$ ratio from its intrinsic value,
which is the opposite of what is observed.

\subsection{Broad Lines}

None of the eight HzPRGs at $z = 2.2-2.6$ were found to have broad
emission-line cores (see Figure 1), nor do they appear to have broad
emission-line wings.  However, it is entirely possible that the
cosmological distances of these galaxies may simply limit the ability
to detect such a feature with 2--4 meter-class telescopes.  The
emission-line spectrum of the lower redshift HzPRG 3C{\ts}22 suggests
that this may indeed be the case.

Figure 2 shows that the H$\alpha$+[N II]{\ts}$\lambda \lambda$6548,
6583 complex for 3C{\ts}22 has a line core width, $\Delta v_{\rm
FWHM} \sim$ 1700 km s$^{-1}$, consistent with the higher redshift
HzPRGs observed (see Table 2).  This emission emanates from extended,
low density gas some distance from the nucleus of the galaxy.
However, note the very broad H$\alpha$ wings ($\Delta v_{\rm FWZI}
\sim$ 7600 km s$^{-1}$) in 3C{\ts}22; such a feature is also visible
in additional data (not shown here) from three KSPEC observing
periods between 1994, July--September, and has also been reported by
Economou et al. (1995) and Rawlings et al. (1995).  Such a feature is
indicative of the presence of a partially obscured Seyfert{\ts}1
nucleus, and is consistent with the hypothesis that FR{\ts}II radio
galaxies are quasars where the broad-line active nucleus is mostly
obscured from the line of sight.  The signal-to-noise ratio of the
spectra of the higher redshift galaxies shown in Figure 1 is simply
insufficient to rule out the presence of a similarly broad component
in the $z = 2.2-2.6$ objects.

It is tempting to speculate that broad line wings, such as those
observed in 3C{\ts}22, may be present in the more distant HzPRGs.
However, current efforts using 2--4 meter-class telescopes suggests
that larger aperture (i.e., 8--10 meter-class telescopes) will be
required to detect broad line wings for HzPRGs at $z > 2$.

\section{Summary}

The following conclusions are drawn from the near-infrared
spectroscopy of a sample of eight HzPRGs at $z = 2.2-2.6$:

\noindent 
{\bf 1.}\ For three of the eight HzPRGs, the emission-line
ratios [O III]{\ts}$\lambda$5007{\ts}/{\ts}H$\beta$ and [S
II]{\ts}$\lambda$6724{\ts}/{\ts}H$\alpha$ are characteristic of
Seyfert{\ts}2 emission, consistent with the spectral type of the
majority of low-redshift, narrow-line, powerful radio galaxies.  The
[O III]{\ts}$\lambda$5007{\ts}/{\ts}H$\beta$ and [S
II]{\ts}$\lambda$6724{\ts}/{\ts}H$\alpha$ ratios and/or limits are
inconclusive for the remaining five  HzPRGs.  Furthermore, of the six
galaxies for which both $H$- and $K$-band spectra have been obtained,
all six  have observed [O
III]{\ts}$\lambda$5007{\ts}/{\ts}(H$\alpha$+[N II] $\lambda
\lambda$6548, 6583) ratios consistent with Seyfert{\ts}2 ionization.

\noindent
{\bf 2.}\  Unlike the emission-line spectra of low-redshift,
narrow-line, powerful radio galaxies, the emission lines of the eight
HzPRGs appear to contribute from $\sim$ 25--85{\ts}\% of the $H$-
and/or $K$-band light from these galaxies.

\noindent
{\bf 3.}\  The southwestern component of TX{\ts}0828+193 shows no
evidence of strong emission lines and comparatively strong continuum
emission, whereas the northeastern component shows strong line
emission and no noticeable continuum emission. If this featureless
component is at the redshift of the radio galaxy, the continuum
emission in TX{\ts0828+193SW may be illumination from an AGN residing
in the NW component. Alternatively, the AGN may reside in the SW
component and thus ionize gas in the NW component, creating the
strong lines observed.  A comparison of these data with low-redshift,
radio galaxies with off-nuclear knots supports the latter
interpretation.

\noindent
{\bf 4.}\  The observed Ly$\alpha${\ts}/{\ts}H$\alpha$ ratios of the
two TX sources are $\sim${\ts}1.5, much less than the value of 16
expected for gas photoionized by a hard nonthermal continuum.  The
discrepancy is most likely due to extinction by dust, though
associated absorption may account for up to a 50\% reduction in
Ly$\alpha$.  The ratio Ly$\alpha${\ts}/{\ts}H$\alpha \sim 1.5$
corresponds to $A_V \sim 0.5-1.0$ (depending on which extinction
curve -- Milky Way, SMC, LMC -- is used), and a corresponding factor
of $\sim${\ts}2 reduction in emission-line and continuum flux.

\noindent
{\bf 5.}\ None of the eight HzPRGs at  $z = 2.2-2.6$ have broad
emission-line cores, and it is not clear whether any have broad
emission-line wings.  However, the near-infrared spectrum of
3C{\ts}22, a radio galaxy at $z = 0.937$ with 1{\ts}$\mu$m luminosity
comparable to that of the eight HzPRGs at $z=2.2-2.6$ and a radio
luminosity only 3--5 times less, shows direct evidence for broad
H$\alpha$ emission wings.  Given that 3C{\ts}22 is at $\sim${\ts}1/3
the luminosity distance of the other galaxies observed, systematic
searches for broad emission-line wings in radio galaxies at $z
>${\ts}2 should be pursued, but may require the use of 8--10
meter-class telescopes.

These new data, along with recent rest-frame UV-to-optical
polarimetry of HzPRGs, are consistent with the idea that many HzPRGs
are predominantly ionized by an active nucleus, and that significant
fraction of their SED may be due to non-thermal emission from the
AGN.

\acknowledgements

It is a pleasure to thank the staffs of the UH 2.2m telescope and the
United Kingdom Infrared Telescope for their generous support during
these observations.  D. Sanders, E. Egami, J. Hora, M. Dickinson, D.
Kim, D. Jewitt, T. Greene, J. Deane, J. Goldader, A. Stockton, J.
Jensen, G. Canalizo, and M. Shepherd provided useful discussions and
assistance, and T. Soifer, L. Armus, and G. Neugebauer provided
photometry from unpublished Keck data.  I am also indebted to D.
Jewitt for providing UKIRT observing time for TX{\ts}0828+193, G.
Knopp for providing the $H$-band image of TX{\ts}0828+193, and J.
Surace for obtaining the $K^{\prime}$-band image of 4C{\ts}40.36.  I
also thank K. Teramura for making final adjustments to the figures
presented in this paper, and the referee, A. Dey, for many useful
comments.  This research was supported in part by NASA grants
NAGS-1741 and NAGW-3938.

\appendix

\section{Near-Infrared Finder Charts}

Much of the difficulty in carrying out a HzPRG spectroscopy program
on 2--4 meter-class telescopes is aligning the galaxy and the slit;
the galaxies are often too faint to see after short integrations with
an imaging array (UH 2.2m) and too faint to ``peak up'' on with the
spectrometer (UKIRT).  With the exception of the KSPEC observations
of 3C{\ts}22, the galaxies were observed by first blind-offsetting
from a star whose position is known to fair accuracy, or by placing
the slit on the sky relative to bright stars in the field of view.
The latter method is preferable to large offsets, and it also makes
it straightforward to check whether or not the galaxy has drifted out
of the slit because the positions of the stars in the field can
constantly be monitored.

To aid in future observations of these galaxies, $K^{\prime}$ images
of eight of the nine galaxies discussed in this paper have been taken
(Figure 5).  Also included in Figure 5 is an $H$-band image of
TX{\ts}0828+193 from Knopp \& Chambers (1997), which, though the
field of view is much smaller than the $K^{\prime}$ images, shows the
distinct double component morphology (see \S 5.3). A summary of the
imaging observations is given in Table 6. All of the $K^{\prime}$
images of the $z > 2$ HzPRGs have been smoothed 4x4 pixels to make it
easier to see the HzPRGs and other faint sources in the field. When
there was some question as to which source was the HzPRG, the radio
coordinates of the galaxy were used to perform astrometry with stars
visible in the STScI Digitized Sky Survey\footnote{The Digitized Sky
Survey were produced at the Space Telescope Science Institute under
the U.S. Government grant NAG W-2166. The images of these surveys are
based on photographic data obtained using the Oschin Schmidt
Telescope on Palomar Mountain and the UK Schmidt Telescope.} images.

\begin{deluxetable}{llllllrrrrlll}
\tablenum{1}
\scriptsize
\tablewidth{0pt}
\tablecaption{Journal of Spectroscopic Observations}
\tablehead{
\colhead{Source} &
\multicolumn{2}{c}{B1950.0} &
\multicolumn{1}{c}{$z$} &
\multicolumn{1}{c}{Telescope} &
\multicolumn{1}{c}{Instr.} &
\multicolumn{1}{c}{P.A.} & 
\multicolumn{1}{c}{Aperture} &
\multicolumn{1}{c}{${\lambda} \over {\Delta \lambda}$\tablenotemark{a}} &
\multicolumn{1}{c}{samp} &
\multicolumn{1}{c}{Date} &
\multicolumn{1}{c}{band} & 
\colhead{Time\tablenotemark{b}}\nl
\cline{2-3} \nl
\colhead{} & 
\multicolumn{1}{c}{R.A.} & 
\multicolumn{1}{c}{Dec.} &
\multicolumn{1}{c}{} &
\multicolumn{1}{c}{} &
\multicolumn{1}{c}{} &
\multicolumn{1}{c}{} & 
\multicolumn{1}{c}{arcsec$^2$} &
\multicolumn{1}{c}{} & 
\multicolumn{1}{c}{} &
\multicolumn{1}{c}{m/y} &
\multicolumn{1}{c}{} &
\colhead{hrs}
} 
\startdata
TX 0200+015&02:00:08.20&01:34:45.95&2.232&UH 2.2m&KSPEC&133&$0.8\times2.0$&760&1$\times$1&$09/93$&H-K&0.3 \nl
           &       &        &     &UKIRT&CGS4&133&$1.5\times6.0$&860&3$\times$2&$10/94$&K&2.2 \nl
           &       &        &     &UH 2.2m&KSPEC&0&$1.0\times0.7$&650&1$\times$1&$11/95$&H-K&2.5 \nl
B3 0731+438&07:31:41.12&43:50:59.0&2.429&UH 2.2m&KSPEC&0&$0.8\times2.0$&760&1$\times$1&$03/94$&H-K&1.7 \nl
           &       &        &     &UH 2.2m&KSPEC&0&$1.0\times2.0$&620&1$\times$1&$12/94$&H-K&2.1 \nl
           &       &        &     &UKIRT&CGS4&0&$2.5\times4.9$&430&4$\times$2&$10/95$&K&1.2 \nl
TX 0828+193&08:28:01.22&19:23:23.8&2.572&UKIRT&CGS4&44&$1.2\times9.8$&860&4$\times$2&$02/96$&H,K&1.2,2.0 \nl
3C 257&11:20:34.55&05:46:46&2.482&UKIRT&CGS4&0&$2.5\times7.4$&860&4$\times$2&04/96&K&1.8 \nl
MG 1744+18&17:44:55.34&18:22:10.8&2.284&UKIRT&CGS4&56&$1.2\times4.9$&860&4$\times$2&$05/95$&K&1.4 \nl
4C 40.36&18:09:19.42&40:44:38.9&2.270&UKIRT&CGS4&80&$1.5\times6.0$&860&6$\times$1&$08/93$&K\tablenotemark{c}&3.9 \nl
4C 48.48&19:31:40.03&48:05:07.1&2.344&UH 2.2m&KSPEC&0&$0.8\times4.0$&760&1$\times$1&$07/93$&J-H-K&2.6 \nl
        &         &         &     &UKIRT&CGS4&32&$1.5\times6.0$&860&6$\times$1&$08/93$&K&6.9 \nl
4C 23.56&21:05:00.96&23:19:37.7&2.480&UKIRT&CGS4&52&$1.5\times6.0$&860&6$\times$1&$08/93$&K&9.6 \nl
        &         &         &     &UKIRT&CGS4&52&$1.5\times6.0$&860&3$\times$2&$10/94$&H,K&1.2,3.6 \nl
\sidehead{Other:}
3C 22&00:48:04.73&50:55:44.8&0.937&UH 2.2m&KSPEC&0&$1.0\times2.0$&620&1$\times$1&$12/94$&J-H-K&1.2 \nl
\enddata
\tablenotetext{a}{At 2.2{\ts}$\mu$m}
\tablenotetext{b}{On-Source Time}
\tablenotetext{c}{$H$-band observations of 4C{\ts}40.36 were obtained by Iwamuro et al. (1996) using the UH~2.2m telescope equipped with a OH-airglow supression spectrograph.}
\end{deluxetable}

\begin{deluxetable}{llclrrrrr}
\tablenum{2}       
\footnotesize
\tablewidth{0pt}
\tablecaption{Emission-Line Properties}
\tablehead{
\multicolumn{1}{c}{Source} &
\multicolumn{1}{c}{Instr.} &
\multicolumn{1}{c}{Date} &
\multicolumn{1}{c}{Line} &
\multicolumn{1}{c}{$\lambda _{\rm obs}$} & 
\multicolumn{1}{c}{${f(\lambda) \over  f(\rm{H}\alpha+[N II])}$} &
\multicolumn{1}{c}{$f(\lambda)$} &
\multicolumn{1}{c}{FWHM} &
\multicolumn{1}{c}{$S/N$} \nl
\multicolumn{1}{c}{} &
\multicolumn{1}{c}{} &
\multicolumn{1}{c}{m/y} &
\multicolumn{1}{c}{} &
\multicolumn{1}{c}{(\micron)} &
\multicolumn{1}{c}{} &
\multicolumn{1}{c}{(Watts m$^{-2}$)} &
\multicolumn{1}{c}{(km s$^{-1}$)} &
\multicolumn{1}{c}{} 
}
\startdata
TX 0200+015&KSPEC&09/93&[O III]$\lambda$5007&1.6183&1.1&\nodata &1080&4.2\nl
           &     &     &H$\alpha$+[N II]&2.1215&1.0&\nodata &1040&3.6\nl
           &CGS4&10/94&H$\alpha$+[N II]&2.1211&1.0&$2.0\times10^{-18}$&1800&15\nl
           &    &     &[S II]$\lambda$6724&2.1715&0.16&$3.3\times10^{-19}$&1200&4.8\nl
           &KSPEC&11/95&[O III]$\lambda$5007&1.6195&1.2&($2.4\times10^{-18}$)&860&5.6\nl
           &     &     &H$\alpha$+[N II]&2.1221&1.0&\nodata &2000&5.9\nl\nl
B3 0731+438&KSPEC&3$-$12/94&[O III]$\lambda$5007&1.7191&2.7&\nodata &680&8.1\nl
           &     &     &[O I]$\lambda$6300&\nodata&$<$0.19&\nodata &(1000)&\nodata \nl
           &     &     &H$\alpha$+[N II]&2.2542&1.0&\nodata &870&5.1\nl
           &     &     &[O I]$\lambda$6300&\nodata&$<$0.08&\nodata &(1000)&\nodata \nl
           &CGS4&10/95&H$\alpha$+[N II]&2.2489&1.0&\nodata &1600&21\nl
           &    &     &[S II]$\lambda$6724&2.3024&0.27&\nodata &1180&12\nl\nl
TX 0828+193&CGS4&02/96&H$\beta$&\nodata&$<$0.36&$<6.5\times10^{-19}$&(1000)&\nodata \nl
           &    &     &[O III]$\lambda$5007&1.7885&1.4\tablenotemark{a}&$2.5\times10^{-18}$&1300&6.9\nl
           &    &     &H$\alpha$+[N II]&2.3463&1.0&$1.8\times10^{-18}$&800&5.0\nl
           &    &     &[S II]$\lambda$6724&\nodata&$<$0.23&$<4.2\times10^{-19}$&(1000)&\nodata \nl\nl
3C 257     &CGS4&04/96&[O I]$\lambda$6300&\nodata&$<$0.11&$<3.0\times10^{-19}$&(1000)&\nodata \nl
           &    &     &H$\alpha$+[N II]&2.2857&1.0&$2.6\times10^{-18}$&1750&6.9 \nl
           &    &     &[S II]$\lambda$6724&2.3429&0.20&$5.2\times10^{-19}$&1180&3.2 \nl\nl
MG 1744+18&CGS4&05/95&H$\alpha$+[N II]&2.1552&1.0&\nodata &1750&7.1\nl
          &    &     &[S II]$\lambda$6724&\nodata&$<$0.14&\nodata &(1200)&\nodata \nl\nl
4C 40.36&CGS4&08/93&[O I]$\lambda$6300&\nodata&$<$0.15&$<5.2\times10^{-19}$&(1000)&\nodata \nl
        &    &     &H$\alpha$+[N II]&2.1462&1.0&$3.4\times10^{-18}$&1660&7.1\nl
        &    &     &[S II]$\lambda$6724&2.1945&0.35&$1.2\times10^{-18}$&1320&4.6\nl\nl
4C 48.48&KSPEC&07/93&[O II]$\lambda$3727&1.2462&0.37&\nodata &1130&3.1\nl
        &     &     &H$\beta$&\nodata&$<$0.26&\nodata &(1000)&\nodata \nl
        &     &     &[O III]$\lambda$5007&1.6745&1.8&($6.0\times10^{-18}$)&640&10\nl
        &     &     &[O I]$\lambda$6300&\nodata&$<$0.12&\nodata &(1000)&\nodata \nl
        &     &     &H$\alpha$+[N II]&2.1945&1.0&\nodata &1240&7.7\nl
        &     &     &[S II]$\lambda$6724&2.2501&$<$0.12&\nodata &(1200)&\nodata \nl
        &CGS4&08/93&[O I]$\lambda$6300&\nodata&$<$0.05&$<1.8\times10^{-19}$&(1000)&\nodata \nl
        &    &     &H$\alpha$+[N II]&2.1920&1.0&$3.5\times10^{-18}$&1110&18 \nl
        &    &     &[S II]$\lambda$6724&2.2440&0.26&$9.2\times10^{-19}$&1240&6.2\nl\nl
4C 23.56&CGS4&08/93&H$\alpha$+[N II]&2.2830&1.0&$1.2\times10^{-18}$&1520&5.9\nl
        &    &10/94&H$\beta$&\nodata&$<$1.7&$<5.9\times10^{-18}$&(1000)&\nodata \nl
        &    &     &[O III]$\lambda$5007&1.7442&1.3\tablenotemark{a}&$4.4\times10^{-18}$&420&5.0\nl
        &    &     &[O I]$\lambda$6300&\nodata&$<$0.09&$<3.2\times10^{-19}$&(1000)&\nodata \nl
        &    &     &H$\alpha$+[N II]&2.2861&1.0&$3.4\times10^{-18}$&1280&6.8\nl
        &    &     &[S II]$\lambda$6724&\nodata&$<$0.14&$<4.6\times10^{-19}$&(1200)&\nodata \nl
\sidehead{Other:}
3C 22      &KSPEC&12/94&H$\beta$&\nodata&$<$0.07&\nodata &(1000)&\nodata \nl
           &     &     &[O III]$\lambda$5007&0.9696&0.76&\nodata &940&5.6 \nl
           &     &     &H$\alpha$+[N II]&1.2712\tablenotemark{b}&1.0&\nodata &1730&16 \nl
           &     &     &                &1.2746\tablenotemark{c}&   & &    &     \nl
           &     &     &[S II]$\lambda$6724&1.3015&0.07&\nodata &880&4.0 \nl
           &     &     &[O I]$\lambda$6300& &$<$0.02&\nodata &(1000)&\nodata \nl
           &     &     &He I$\lambda$10830&2.0996&0.15&\nodata &1660&8.6 \nl
\enddata
\tablenotetext{a}{For TX 0828+193 and 4C 23.56, [O III] and H$\alpha$+[N II]
were not observed simultaneously, so the relative fluxes may be affected by
blind offset errors.}
\tablenotetext{b}{Observed wavelength of H$\alpha$$\lambda$6563.}
\tablenotetext{c}{Observed wavelength of [N II]$\lambda$6583.}
\tablecomments{All upper limits are 3$\sigma$ based on the $rms$ of
the given spectrum and the adopted FWHM (in parentheses).
[O III] fluxes in parentheses are adopted values 
based on the CGS4 H$\alpha$ flux for the radio galaxy and
$f({\rm [O III]} \lambda 5007)/
f({\rm H}\alpha+[{\rm N II}])$ measured with KSPEC.}
\end{deluxetable}

\begin{deluxetable}{lrrrrr}
\tablenum{3}         
\tablewidth{0pt}
\tablecaption{Observed Emission-Line Ratios of HzPRGs \& Low-z Galaxies}
\tablehead{
\colhead{Source} &
\multicolumn{1}{c}{${{\rm [O III]}\lambda5007 \over {\rm H}\alpha+{\rm [N II]}}$} &
\multicolumn{1}{c}{${{\rm [S II]}\lambda6724  \over {\rm H}\alpha+{\rm [N II]}}$} &
\multicolumn{1}{c}{${{\rm H}\alpha \over {\rm [S II]}\lambda6724}$} &
\multicolumn{1}{c}{${{\rm H}\alpha \over {\rm [N II]}\lambda \lambda6548+6583}$} &
\multicolumn{1}{c}{${{\rm [N II]}\lambda \lambda6548+6583 \over {\rm [S II]}\lambda6724}$} \nl}
\startdata
HzPRG & $1.6\pm0.6$ & $0.25\pm0.07$ &\nodata&\nodata&\nodata   \nl
LzPRG & $0.99\pm0.34$ & $0.27\pm0.06$ & $1.6\pm0.6$ & $0.74\pm0.22$ & $2.2\pm0.4$ \nl
Seyfert2 & $0.80\pm0.40$ & $0.18\pm0.07$ & $0.98\pm0.45$ & $3.0\pm1.0$ & $3.5\pm
1.7$ \nl
LINER & $0.062\pm0.044$ & $0.25\pm0.08$ & $0.71\pm0.25$ & $1.8\pm0.6$ & $2.6\pm0.7$ \nl
%Seyfert2 & $0.80\pm0.40$ & $0.18\pm0.07$ & $0.98\pm0.45$ & $3.0\pm1.0$ & $3.5\pm1.7$ \nl
HII    & $0.068\pm0.080$ & $0.17\pm0.04$ & $1.7\pm0.4$ & $3.9\pm0.8$ & $2.4\pm0.7$ \nl
\enddata
\end{deluxetable}

\begin{deluxetable}{lcll}
\tablenum{4}         
\tablewidth{0pt}
\tablecaption{Emission-Line Contribution to Infrared Flux}
\tablehead{
\colhead{Source} &
\multicolumn{1}{c}{Band} &
\multicolumn{1}{c}{Magnitude\tablenotemark{a}} & 
\multicolumn{1}{c}{Line/Total\tablenotemark{b}} \nl
\multicolumn{3}{c}{} &
\multicolumn{1}{c}{(\%)} \nl
}
\startdata
TX 0200+015  & $H$ & 18.90$\pm0.05$  & 25$\pm2$  \nl
TX 0200+015   & $K$ & 18.27$\pm0.03$ & 28$\pm2$  \nl
4C 40.36      & $K$ & 18.01$\pm0.07$ & 44$\pm6$ \nl
4C 23.56      & $H$ & 19.73$\pm0.10$   & 98$\pm21$ \nl
              & $K$ & 18.95$\pm0.07$   & 77$\pm13$  \nl
\tablenotetext{a}{These broad-band 
magnitudes are determined with the same apertures
used for the spectroscopic observations (Table 1).}
\tablenotetext{b}{Calculation of line contribution to the broad-band
magnitude of the HzPRGs, where the magnitudes are determined
from imaging data by L. Armus (private communication).}
\enddata
\end{deluxetable}

\begin{deluxetable}{lcrrrrrr}
\tablenum{5}
\tablewidth{6in}
\tablecaption{Color-Excess Calculations}
\tablehead{
\multicolumn{1}{c}{Source} &
\multicolumn{1}{c}{$\frac{{\rm [N~II]}}{{\rm H}\alpha}$\tablenotemark{a}} &
\multicolumn{2}{c}{$\frac{{\rm Ly}\alpha}{{\rm H}\alpha}$\tablenotemark{b}} &
\multicolumn{1}{c}{} &
\multicolumn{3}{c}{Intrinsic {\it E(B--V)}\tablenotemark{c}} \nl
\cline{3-4} \cline{6-8} \nl
\multicolumn{1}{c}{} &
\multicolumn{1}{c}{} &
\multicolumn{1}{c}{obs} &
\multicolumn{1}{c}{corr} &
\multicolumn{1}{c}{} &
\multicolumn{1}{c}{MW} &
\multicolumn{1}{c}{LMC} &
\multicolumn{1}{c}{SMC} 
}
\startdata
TX 0200+015 & 0.19 & 1.0$\pm$0.1 & 1.0$\pm$0.1 & & 0.38$\pm$0.02 &0.26$\pm$0.02&0.18$\pm$0.01 \nl
            & 1.00 & 1.7$\pm$0.2 & 1.7$\pm$0.2 & & 0.31$\pm$0.02 &0.21$\pm$0.02 &0.14$\pm$0.01 \nl
            & 3.74 & 4.0$\pm$0.6 & 4.0$\pm$0.6 & & 0.19$\pm$0.02 &0.13$\pm$0.01 &0.09$\pm$0.01 \nl
TX 0828+193 & 0.19 & 0.86$\pm$0.19 & 0.95$\pm$0.2 & & 0.39$\pm$0.03 &0.27$\pm$0.02 &0.18$\pm$0.02 \nl 
            & 1.00 & 1.4$\pm$0.3 & 1.6$\pm$0.3 & & 0.32$\pm$0.03 &0.22$\pm$0.02 &0.15$\pm$0.01 \nl
            & 3.74 & 3.4$\pm$0.7 & 3.8$\pm$0.8 & & 0.20$\pm$0.03 &0.14$\pm$0.02 &0.09$\pm$0.02 \nl
\enddata
\tablenotetext{a}{1.0 = most probable value (Eales \& Rawlings 1993), 
 0.19--3.74 = extreme values}
\tablenotetext{b}{Ly$\alpha$/H$\alpha$ observed and corrected for Galactic extinction.}
\tablenotetext{c}{Calculated color-excess assuming a Milky Way, LMC, and SMC extinction curve 
and corrected Ly$\alpha$/H$\alpha$.}
\end{deluxetable}

\begin{deluxetable}{lllcccc}
\tablenum{6}
\tablewidth{0pt}
\tablecaption{Journal of Imaging Observations}
\tablehead{
\colhead{Source} &
\multicolumn{1}{c}{Telescope} &
\multicolumn{1}{c}{Instr.} &
\multicolumn{1}{c}{Pixel Scale} &
\multicolumn{1}{c}{Band} &
\multicolumn{1}{c}{Date} &
\colhead{Time}\nl
\colhead{} & 
\multicolumn{1}{c}{} &
\multicolumn{1}{c}{} &
\multicolumn{1}{c}{arcsec/pixel} &
\multicolumn{1}{c}{} &
\multicolumn{1}{c}{m/y} &
\colhead{min.}
} 
\startdata
TX 0200+015&UH 2.2m&QUIRC1024x1024&0.1886&K$^{\prime}$&$12/95$& 106 \nl
B3 0731+438&UH 2.2m&QUIRC1024x1024&0.1886&K$^{\prime}$&$12/95$& 40\nl
TX 0828+193\tablenotemark{a}&UKIRT&IRCAM3&0.286&H&$03/96$& 160 \nl
3C 257&UH 2.2m&QUIRC1024x1024&0.1886&K$^{\prime}$&$04/96$& 20 \nl
MG 1744+18&UH 2.2m&QUIRC1024x1024&0.1886&K$^{\prime}$&$07/97$& 45 \nl
4C 40.36&UH 2.2m&QUIRC1024x1024&0.06084&K$^{\prime}$&$07/97$& 56 \nl
4C 48.48&UH 2.2m&QUIRC1024x1024&0.1886&K$^{\prime}$&$12/95$& 28 \nl
4C 23.56&UH 2.2m&QUIRC1024x1024&0.1886&K$^{\prime}$&$12/95$& 40 \nl
3C 22&UH 2.2m&QUIRC1024x1024&0.1886&K$^{\prime}$&$12/95$& 18 \nl
\enddata
\tablenotetext{a}{The $H$-band image of TX0828+193 is taken from Knopp \& Chambers (1997).}
\end{deluxetable}

\vfill\eject

\centerline{Figure Captions}

\vskip 0.3in

\noindent
Figure 1. Rest-frame optical spectroscopy for the sample of 8 HzPRGs
at redshifts 2.2 to 2.6.\ \ (a) TX{\ts}0200+015;  (b)
B3{\ts}0731+438; (c) TX{\ts}0828+193; (d) 3C{\ts}257, MG{\ts}1744+18,
4C{\ts}40.36; (e) 4C{\ts}48.48; (f) 4C{\ts}23.56. The resolutions and
the samplings of all of the observations are listed in Table 1. For
the KSPEC observations, the pixel sampling for $J, H,$ and $K$-bands
are 12, 14, and 20 $\mu$m pixel$^{-1}$, respectively.  For the CGS4
observations, the pixel samplings are as follows:  6.6 $\mu$m
pixel$^{-1}$ for B3{\ts}0731+438, MG{\ts}1744+18, and
TX{\ts}0828+193; 5.6 $\mu$m pixel$^{-1}$ for 4C{\ts}48.48,
4C{\ts}40.36, and the 1993, August observations of 4C{\ts}23.56; 11
$\mu$m pixel$^{-1}$ for TX{\ts}0200+015 and the 1994, October
observations of 4C{\ts}23.56.  The CGS4 spectra of TX{\ts}0200+015,
4C{\ts}40.36, and TX{\ts}0828+193SW ($K$ band) have been smoothed an
additional 10 pixels , 10 pixels, and 100 pixels, respectively.

\noindent
Figure 2.  Rest-frame 4800--10840 \AA$~$spectroscopy of the radio
galaxy 3C{\ts}22. The pixel sampling for $J, H,$ and $K$-bands are
12, 14, and 20 $\mu$m pixel$^{-1}$, respectively.  Note the broad
H$\alpha$ wings.

\noindent
Figure 3.  log ( [O III]{\ts}$\lambda$5007{\ts}/{\ts}H$\beta$ ) vs.
log ( [S II]{\ts}$\lambda$6724{\ts}/{\ts}H$\alpha$ ) for 8 HzPRGs.
For all of the data, the ratios are plotted assuming the lower limit
value [N II]{\ts}$\lambda$6583{\ts}/{\ts}H$\alpha$ = 0.19.  The two
additional [N II]{\ts}$\lambda$6583{\ts}/{\ts}H$\alpha$ ratios of 1.0
and 3.74 (see text) are indicated by the small solid circles on the
thin lines.

\noindent
Figure 4.  a) $L_{\rm [O III]{\ts}\lambda \lambda4959, 5007}$ vs.
$P_{151 {\rm MHz}}$, adapted from Eales \& Rawlings (1996).
Symbols---($\bullet$) sources from this paper, (+) objects in Eales
\& Rawlings (1996).  The radio data for the TX galaxies discussed in
this paper are determined from data in R\"{o}ttgering (1993), and the
radio data for the 4C sources and B3{\ts}0731+438 are determined from
data in the NASA/IPAC Extragalactic Database (NED).  The [O
III]{\ts}$\lambda \lambda$4959, 5007 luminosities of TX{\ts}0200+015,
B3{\ts}0731+438, and 4C{\ts}48.48 have been determined using KSPEC
observations of the [O III]{\ts}$\lambda \lambda$4959,
5007{\ts}/{\ts}H$\alpha$+[N II]{\ts}$\lambda \lambda$6548, 6583
ratios and CGS4 measurements of H$\alpha$+[N II]{\ts}$\lambda
\lambda$6548, 6583 luminosities (the H$\alpha$+[N II]{\ts}$\lambda
\lambda$6548, 6583 luminosity of B3{\ts}0731+438 is the measured
value from Eales \& Rawlings 1993).  b) $L_{\rm [O III]{\ts}\lambda
\lambda4959, 5007}${\ts}/{\ts}$P_{151 {\rm MHz}}$ vs.  $P_{151 {\rm
MHz}}$.

\noindent
Figure 5. Near-infrared images of the nine HzPRGs discussed in the
paper.  A scale bar is provided in each image, and the identification
is marked with two ticks.  In all images, North is up, and East is to
the left.

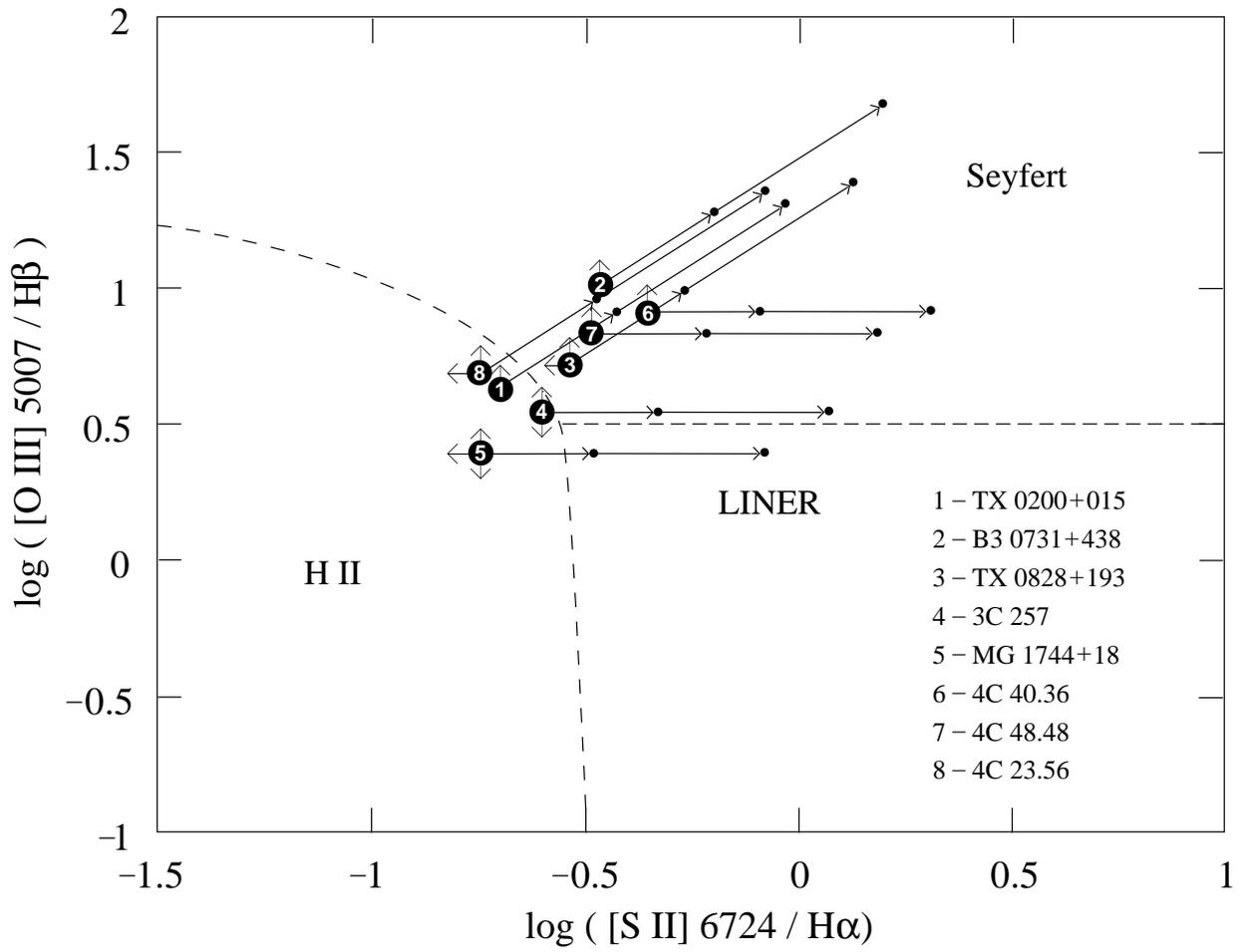
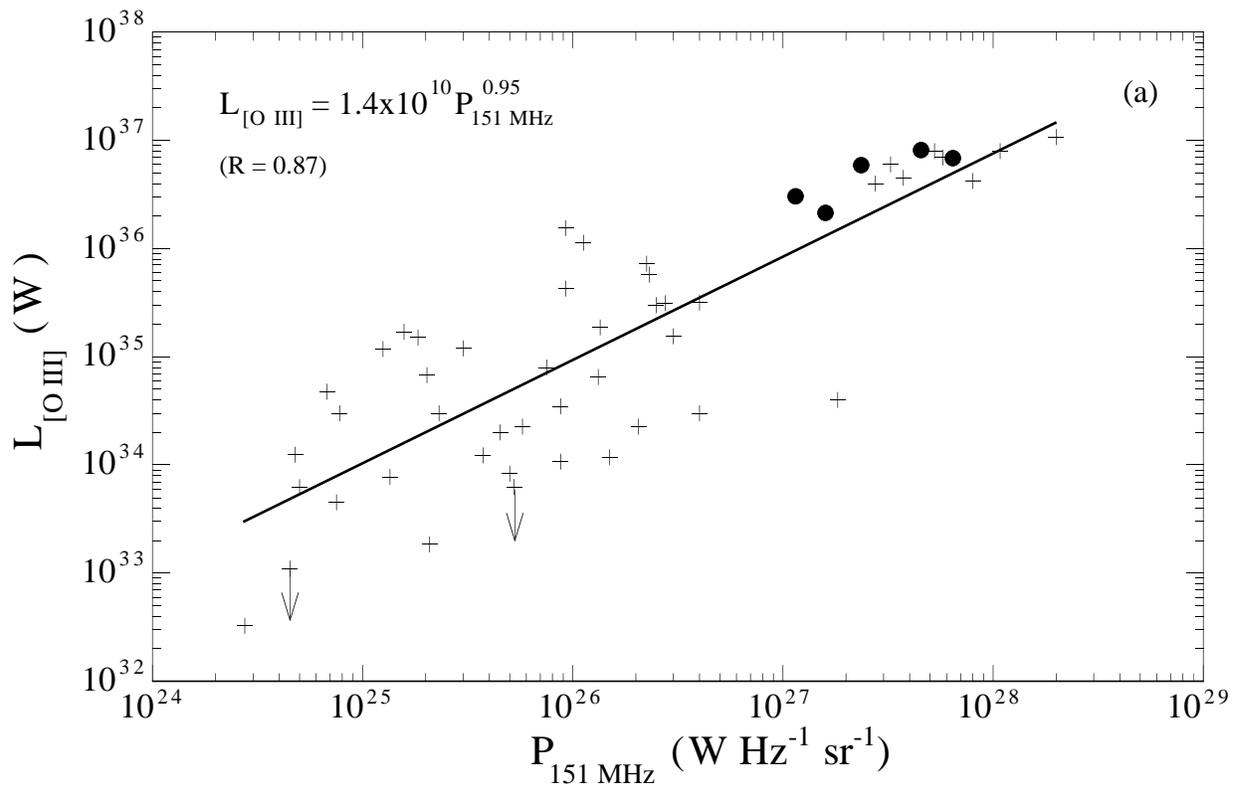
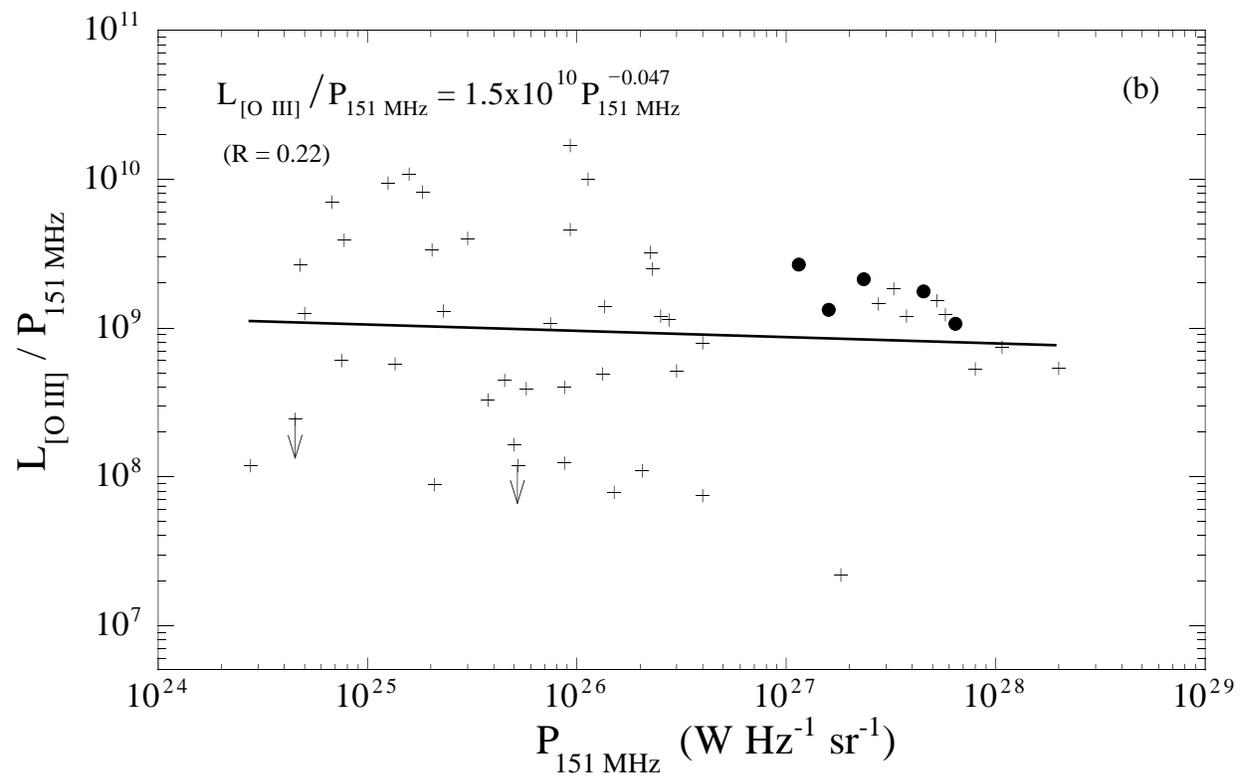
\end{document}